\documentclass{article}
\usepackage{spconfa4,amsmath,graphicx}
\usepackage{amssymb,amsfonts,multirow}

\title{Neural Array-Generic Direction-of-Arrival Estimation Exploiting Array Transfer Functions}
\name{Mikko Heikkinen$^{\dag}$ \qquad Archontis Politis$^{\star}$ \qquad Konstantinos Drossos$^{\ddag}$ \qquad Tuomas Virtanen$^{\star}$}
\address{$^{\dag}$$^{\ddag}$ Nokia, Tampere$^{\dag}$ / Espoo$^{\ddag}$, Finland \\
    $^{\star}$ Tampere University, Tampere, Finland}
\begin{document}
%
\maketitle
\begin{abstract}
Direction-of-arrival (DoA) estimation is a key component of multichannel audio processing, yet many deep learning approaches remain tied to the microphone arrays used during training and generalize poorly to unseen devices. This paper proposes an array-generic neural DoA estimation framework using measured or simulated complex directional array transfer functions (ATFs) matched to real-world multi-microphone devices. The method processes multichannel spectrograms and ATF metadata with separate convolutional encoders, fuses the resulting representations through cross-attention, and predicts source directions using a multi-source Cartesian vector output formulation. Experiments on simulated 2D and 3D localization tasks under reverberation and diffuse babble noise show that the proposed approach generalizes to previously unseen arrays, including mobile-phone-like configurations, without major performance degradation, while remaining competitive with conventional and learning-based baselines.
\end{abstract}
\section{Introduction}
\label{sec:intro}

Direction-of-arrival (DoA) estimation is a core task in micro\-phone-array (MA) signal processing, where the objective is to infer the direction of one or more acoustic sources from multichannel recordings. It plays an important role in a wide range of applications, including beamforming, source separation and enhancement, speaker tracking, robot audition, and spatially aware human–machine interfaces. Reliable DoA estimates are often essential for these downstream systems, as their performance depends directly on the quality of the spatial information provided.

Classical approaches to DoA estimation include time-difference-of-arrival (TDOA)-based methods, often using generalized cross-correlation with phase transform (GCC-PHAT)~\cite{knapp2003generalized} to estimate inter-microphone delays that can then be mapped to source direction. Steered response methods, such as SRP-PHAT~\cite{dibiase2000high}, search over candidate directions by steering a beamformer and evaluating its output power. MUSIC~\cite{schmidt1986multiple} is a subspace method that exploits the eigenstructure of the spatial covariance matrix to form a high-resolution spatial pseudo-spectrum. These methods are widely used and well understood, but their performance can deteriorate in the presence of noise, reverberation, interfering sources, or array mismatch. In particular, subspace- and steering-based approaches rely on sufficiently accurate array calibration and propagation modeling.

More recently, deep neural network (DNN)-based DoA estimation methods have shown strong potential for improved robustness and accuracy under reverberation and noise \cite{grumiaux2022survey}. However, nearly all such methods are matched to the MA used during training, making deployment on unseen arrays difficult and often requiring costly retraining. To address this limitation, a few works have incorporated array metadata to improve generalization. For example, \cite{schwartz2023array} uses SRP-PHAT maps with a U-Net architecture and demonstrates robustness to limited array perturbations, though not full array generalisation. FCGA \cite{kowalk2023geometry} combines GCC-PHAT features with microphone coordinates for azimuth estimation across planar omnidirectional arrays. Neural-SRP \cite{grinstein2023neural} uses GCC-based features and microphone coordinates to produce DoA maps for azimuth and elevation estimation, and GI-DOAEnet \cite{baek2025dnn} combines latent signal features with positional embeddings from microphone coordinates and introduces a curriculum training strategy. PhaseCoder \cite{dementyev2026phasecoder} proposes an array-generic spatial audio encoder using multichannel STFTs and microphone coordinates, with learned spatial embeddings evaluated on downstream tasks including DoA estimation. Together, these studies show that array-aware learning can improve generalization beyond a single fixed array.

However, geometry-based metadata fully describes only omnidirectional arrays under free-field propagation. This assumption often fails in practical consumer devices, such as mobile phones, smart glasses, head-mounted displays, and robotic platforms where compact and irregular MAs are shaped by industrial design constraints. In such cases, microphone coordinates alone cannot model the full acoustical effects of the array to sounds incident from various directions, including frequency-dependent microphone directivities, and device-body scattering effects. Therefore, geometry-only methods remain limited for realistic array-generic DoA estimation.

In this paper, we address this limitation by replacing geometry-only metadata with complex directional acoustic transfer functions (ATFs). Using ATFs as array descriptors allows the method to generalize to unseen arrays with the same number of microphones while also supporting arrays with complex directivities and scattering effects, extending beyond the omnidirectional free-field setting of prior work.

We propose a neural DoA estimation framework that takes both multichannel signals and ATF metadata as input. Separate convolutional encoders process the signal observations and ATF representation, after which cross-attention-based fusion and a shared decoder produce the final DoA estimates. This enables joint learning of signal features and array characteristics while explicitly conditioning the estimator on array-specific transfer behavior.

The proposed method is evaluated against both conventional and DNN-based approaches using simulated data with reverberation and diffuse babble noise. Experiments consider both a single-source azimuth-only task on planar omnidirectional arrays and a more challenging multiple-source task involving omnidirectional and mobile-phone-style four-microphone arrays. Results show that the proposed method learns the DoA estimation task effectively, scales well with the number of sources, and generalizes to unseen arrays without significant performance loss relative to models trained and evaluated on a single fixed array.

\section{Proposed Method}
\label{sec:format}

\subsection{Signal model and task}
\label{ssec:signal_model}

We consider a multichannel time--frequency signal recorded by a MA. Let $Y_m(\omega,t)$ denote the observation at microphone $m$, angular frequency index $\omega$, and time frame $t$. The signal is modeled as
\begin{align}
Y_m(\omega,t)=&\sum_{s\in\mathcal{S}_t}\left(X^{\mathrm{d}}_{s,m}(\omega,t)+X^{\mathrm{r}}_{s,m}(\omega,t)\right)+\\
&+V_m(\omega,t),\nonumber
\end{align}
where $\mathcal{S}_t$ is the set of active sources, $X^{\mathrm{d}}_{s,m}(\omega,t)$ and $X^{\mathrm{r}}_{s,m}(\omega,t)$ are the direct-path and reverberant components of source $s$, respectively, and $V_m(\omega,t)$ is additive noise. The direct-path component is written as
\begin{equation}
X^{\mathrm{d}}_{s,m}(\omega,t)=X_s(\omega,t,\mathbf{u}_{s})\,H_{s,m}(\omega,\mathbf{u}_{s})\,D_{s}(\omega,r_{s}),
\end{equation}
where $X_s(\omega,t,\mathbf{u}_{s})$ is the source signal, $H_{s,m}(\omega,\mathbf{u}_{s})$ is the directional array transfer function, $\mathbf{u}_{s}$ is a unit vector denoting the DoA of static source $s$ at array center, and $D_{s}(\omega,r_{s})$ models propagation effects over distance $r_{s}$.

$H_{s,m}(\omega,\mathbf{u}_{s})$ provides localization-relevant spatial information that can support generalization to unseen arrays. The task is therefore to estimate the source direction(s) of arrival (DoAs) from the multichannel signal representation $\mathbf{X}$ together with the ATFs $\mathbf{H}$.

\subsection{DNN Model}
\label{ssec:dnn_model}

The proposed model is adapted from the array-generic Ambisonics encoder presented in \cite{heikkinen2026beyond}. In particular, the signal and directivity encoders follow the design, with the modifications detailed below. The Ambisonics decoder is replaced with a decoder for DoA estimation. All two-dimensional convolutions are replaced with CoordConv \cite{liu2018intriguing} to better capture location-dependent frequency patterns relevant to DoA estimation.

For an array with $M$ microphones, the complex multichannel spectrogram is converted to a real-valued tensor by concatenating its real and imaginary parts, yielding $\mathbf{X}\in\mathbb{R}^{2M\times F\times T}$, where $F$ and $T$ denote the numbers of frequency bins and time frames. The array-transfer-function feature is constructed analogously from directional transfer functions over $D$ sampled directions, yielding $\mathbf{H}\in\mathbb{R}^{2M\times D\times F}$.

The signal encoder processes $\mathbf{X}$ along $T$ and $F$ using three 2D convolutional layers with $3\times 3$ kernels and output channel dimensions $32$, $64$, and $64$. The first two layers use group normalization with $8$ channels per group, whereas the last layer uses instance normalization. The resulting latent representation is $\mathbf{Z}_X\in\mathbb{R}^{C\times T\times E}$, where $C$ denotes the number of feature channels and $E$ the frequency embedding dimension.

The directivity encoder processes $\mathbf{H}$ using two convolutional layers with kernel size $3\times 1$, such that the convolutions operate only along the frequency dimension and do not mix information across directions $D$. The two layers have $32$ and $64$ output channels, respectively. The first layer uses group normalization with $8$ channels per group, whereas the second layer uses layer normalization with normalized shape $[D,E]$. The resulting latent representation is $\mathbf{Z}_H\in\mathbb{R}^{C\times D\times E}$.

Latent features $\mathbf{Z}_X$ and $\mathbf{Z}_H$ are fused by multi-head cross-attention. For each feature channel, queries are derived from $\mathbf{Z}_X$, while keys and values are derived from $\mathbf{Z}_H$. The same attention weights are shared across all feature channels $C$. Attention is computed as in \cite{vaswani2017attention}, producing a fused representation $\mathbf{Z}_{\mathrm{Attn}}\in\mathbb{R}^{C\times T\times E}$. This conditions signal-derived spatial cues on the array-dependent directional characteristics.

The decoder first applies a $1\times 1$ Conv2D layer to reduce the number of feature channels to one. It then uses 2D average pooling with kernel size $[20,1]$ to decimate the temporal dimension while preserving the embedding dimension. The resulting sequence is processed by a two-layer MLP along the embedding dimension, with layer dimensions $E$ and $3S$, where $S$ is the maximum number of simultaneous sources. The decoder output is therefore $\mathbf{Y}_{\mathrm{DOA}}\in\mathbb{R}^{T_{\mathrm{out}}\times S\times 3}$, where $T_{\mathrm{out}}$ denotes the number of output frames after temporal decimation and the last dimension contains Cartesian direction vectors.

At each output frame, the model predicts $S$ Cartesian DoA vectors, while the ground truth (GT) consists of up to $S$ Cartesian DoA vectors. When fewer than $S$ GTs are available, a GT vector duplication strategy similar to multi-ACCDOA~\cite{shimada2022multi} is used to match the fixed output cardinality. Training employs a permutation-invariant loss based on Cartesian distances between all predictions and GTs. The optimal assignment between predictions and GTs, corresponding to Hungarian matching in PIT, is implemented using the differentiable SinkPIT loss approximation~\cite{tachibana2021towards}. During inference, duplicate predictions are merged by clustering.

\section{Evaluation}
\label{sec:evaluation}

We evaluate the proposed method in two simulated localization scenarios, both involving reverberant rooms and diffuse babble noise. The first scenario considers 2D localization with a single source and planar omnidirectional arrays, where only azimuth is estimated. The second scenario extends the task to 3D localization with up to four simultaneous sources, including both free-field and mobile-phone-like arrays, and evaluates both azimuth and elevation. Additional 3D variants assess robustness to reverberation via anechoic conditions and array generalization via single-array training.

\subsection{Baselines}
We implemented and compared against two baselines. The first is the fully connected geometry-aware (FCGA) method~\cite{kowalk2023geometry}, a DNN-based approach that takes GCC-PHAT maxima and microphone coordinates as input and formulates azimuth estimation as classification over discrete directions. The second is MUSIC, a reference subspace localization method that uses array signals together with array ATFs to form a pseudo-spectrum. For multi-source scenarios using MUSIC, we adopt the iterative peak suppression procedure of ~\cite{mccormack2023spatial}.

\subsection{Datasets}
The 2D dataset follows the FCGA simulation setup. Planar arrays are sampled within a $0.4 \times 0.4$~m area and placed in reverberant rooms generated with the image-source method~\cite{scheibler2018pyroomacoustics}. A single source is placed at the same elevation as the array, with azimuth quantized at $5^\circ$ resolution. Room dimensions vary between 4--6~m in width, 8--10~m in depth, and 2.5--3.5~m in height, with $RT_{60}$ between 0.23~s and 0.8~s. Source distance is 1--3~m. Diffuse babble noise is generated by placing multiple decorrelated source instances on a spherical grid around the array. The data are simulated as source-to-microphone impulse responses, while the source signal material is selected separately during training. The 2D data use an 8~kHz sampling rate, and 50\% of source signals are replaced by white noise as in~\cite{kowalk2023geometry}. Babble-noise SNR varies between 0--30~dB in training and is fixed to 20~dB in evaluation.

The 3D dataset extends the same simulation procedure to varying source elevations and 1--4 simultaneous sources, with a minimum angular separation of $10^\circ$. MAs have four microphones. Free-field arrays are sampled within a $0.4 \times 0.4 \times 0.4$~m volume. In addition, mobile-phone-like arrays are simulated using boundary-element method on cuboid device meshes~\cite{betcke2021bempp}. These consist of four microphones with constrained pairwise spacing, with 1100 configurations used for training and validation and 300 for testing. We also consider an anechoic 3D variant using only direct sound, and a single-array variant in which all examples use the same mobile-phone-like array. For each dataset variant, $200\,000$, $20\,000$, and $20\,000$ impulse-response examples are generated for training, validation, and testing, respectively. The single-array dataset is one fifth of this size. The 3D data use a 16~kHz sampling rate.

In both the 2D and 3D cases, the simulations define the acoustic transfer paths, whereas source content is sampled dynamically. At each training step, speech and noise signals are randomly selected and convolved with the precomputed impulse responses to generate the final multichannel mixtures. Speech signals are drawn from LibriSpeech~\cite{panayotov2015librispeech}, and babble noise is taken from WHAM!~\cite{Wichern2019WHAM}, following the training, validation, and test splits of the respective datasets.

\begin{table}[t]
\centering
\caption{2D DoA estimation results for single-source azimuth estimation. $\text{F}_{1}$-score at five degrees ($\text{F}_{1}$@${5^\circ}$) is higher-better, while localization error (LE) is lower-better.}
\label{tab:results_2d}
\bigskip
\begin{tabular}{lcc}
Method & $\text{F}_{1}$@${5^\circ}$ & LE ($^\circ$) \\
\hline
PROPOSED & 0.75 & 4.0 \\
FCGA & 0.81 & 9.8 \\
MUSIC & 0.97 & 1.5 \\
\end{tabular}
\unskip
\end{table}
\subsection{Model and training setup}
Training and evaluation use 1 second STFT spectrograms with a window length and FFT size of 256, a hop size of 128, and Hann windowing. The directional ATF grid size and latent embedding dimension were fixed to $D=924$ and $E=128$. Optimization is performed using Adam with learning rate $3 \times 10^{-4}$. The batch size is 16, with an effective batch size of 64 obtained through gradient accumulation over four steps. Early stopping with a patience of 50 epochs is applied.

\begin{table*}[t]
\centering
\caption{Combined 3D DoA estimation results. ``Rev+noise'' denotes reverberant conditions with babble noise, and ``dry'' denotes direct-sound-only conditions. $\text{F}_{1}$-score ($\text{F}_{1}$) is higher-better, while localization error (LE) is lower-better. MUSIC $\text{F}_{1}$-scores are omitted because, with oracle source counts and a $180^\circ$ threshold, the metric is not meaningful.}
\label{tab:results_3d_all}
\smallskip

\begin{tabular}{lllcccccccc}
\multicolumn{2}{l}{\multirow{2}{*}{Condition}} & \multirow{2}{*}{Method} & \multicolumn{2}{c}{1 source} & \multicolumn{2}{c}{2 sources} & \multicolumn{2}{c}{3 sources} & \multicolumn{2}{c}{4 sources} \\
& & & $\text{F}_{1}$ & LE ($^\circ$) & $\text{F}_{1}$ & LE ($^\circ$) & $\text{F}_{1}$ & LE ($^\circ$) & $\text{F}_{1}$ & LE ($^\circ$) \\
\hline
\multirow{4}{*}{omni} & \multirow{2}{*}{Dry} & PROPOSED & 1.00 & 18.7 & 0.69 & 25.6 & 0.55 & 26.1 & 0.47 & 26.5 \\
 & & MUSIC & - & 5.3  & - & 14.7 & - & 27.9 & - & 54.1 \\
\cline{2-11}
& \multirow{2}{*}{Rev+noise} & PROPOSED & 0.99 & 22.9 & 0.69 & 28.7 & 0.58 & 31.2 & 0.50 & 30.7 \\
& & MUSIC    & - & 8.4  & - & 28.7 & - & 51.1 & - & 53.4 \\
\hline
\multirow{3}{*}{mobile}& \multirow{2}{*}{Rev+noise}        & PROPOSED & 0.99 & 27.4 & 0.72 & 34.3 & 0.61 & 35.5 & 0.54 & 34.1 \\
&         & MUSIC    & - & 6.5  & - & 26.5 & - & 41.8 & - & 53.4 \\
\cline{2-11}
& Rev+noise, single & PROPOSED & 1.00 & 20.6 & 0.78 & 26.2 & 0.73 & 30.0 & 0.67 & 29.9 \\
\end{tabular}
\end{table*}
\subsection{Metrics}
We adopt a subset of the DCASE localization metrics ~\cite{politis2020overview} and use their implementation. Localization error (LE) and $F_1$-score ($F_1$) are used in all evaluations. The $F_1$ has a spatial threshold parameter, which treats estimates that fall outside the threshold as false positives. LE is averaged across all predictions. In the 2D scenario, we report LE and error rate with a spatial threshold of $5^\circ$, matching the FCGA evaluation protocol. In the 3D scenario, where the number of active sources varies, we disable the spatial threshold by setting it to 180$^\circ$ so that the $F_1$ focuses on the number of detected sources.
\section{Results and discussion}\label{sec:results}
Table~\ref{tab:results_2d} summarizes the 2D single-source azimuth estimation results. The $F_1$-score was computed using a spatial threshold of $5^\circ$, meaning that predictions with angular error above $5^\circ$ were counted as false detections, whereas localization error (LE) was computed independently of this threshold.

In the 2D case, MUSIC achieved the best overall performance, with the highest $F_1$ and lowest LE. The proposed method ranked between MUSIC and FCGA and although its $F_1$ was lower than that of FCGA, its LE was substantially smaller. This suggests that FCGA produced larger outliers, even if many predictions remained within the $5^\circ$ threshold. The original FCGA paper reported higher performance on similar evaluation (LE 1.47$^\circ$, and accuracy 96.1\%), which may be explained by the different dataset or implementation detail.  Overall, the proposed method works well in this single source 2D setting and is competitive although it does not surpass MUSIC.

Table~\ref{tab:results_3d_all} combines all 3D evaluations. In the reverberant and noisy case, MUSIC was clearly strongest for a single source, especially in terms of localization accuracy. However, as the number of simultaneous sources increased, its LE grew rapidly, whereas the proposed method degraded more gradually. This indicates that MUSIC remained stronger in the easier cases, while the proposed method showed a more stable trend as scene complexity increased. The comparison should also be interpreted in light of the evaluation setup: MUSIC was given oracle information about the number of active sources, whereas the proposed model had to both detect and localize them.

The direct-sound-only results show that removing reverberation and babble noise improved the proposed method slightly, but the difference was not dramatic. This suggests that the model is reasonably robust to the tested acoustic degradations. At the same time, the gap to MUSIC in the single-source case indicates that the proposed model would likely benefit from improved modeling capacity or a stronger training strategy.

The comparison between the mobile array generalization setting and the mobile single-array setting shows only a moderate performance gap. The same degradation trend with increasing number of sources is visible in both cases, and the single-array model provides only limited improvement. This supports the main motivation of the paper that conditioning on ATFs enables viable array generalization.

Overall, the experiments show that the proposed method learns the DoA estimation task across different array types and acoustic conditions and generalizes to unseen arrays without major performance collapse. While MUSIC remains stronger in single-source scenarios, the proposed method is competitive in multi-source cases, supporting array-generic localization using ATF-based conditioning.

\section{Conclusion}
\label{sec:conclusion}
This paper presented an array-generic neural DoA estimation framework conditioned on ATFs, using separate signal and metadata encoders with cross-attention fusion. Experiments on simulated 2D and 3D tasks showed that the method generalizes to previously unseen arrays without major performance loss. Although MUSIC remained stronger in single-source conditions, the proposed model degraded more gracefully as the number of simultaneous sources increased and achieved viable performance also for mobile-phone-like arrays. Future work will address single-source accuracy, varying microphone counts, and validation on measured and real-world data.

\bibliographystyle{IEEEbib}
\bibliography{refs}

@article{knapp2003generalized,
  title={The generalized correlation method for estimation of time delay},
  author={Knapp, C. and Carter, G.},
  journal={IEEE Transactions on Acoustics, Speech, and Signal processing},
  volume={24},
  number={4},
  pages={320--327},
  year={1976},
  publisher={IEEE}
}

@inproceedings{shimada2022multi,
  title={Multi-accdoa: Localizing and detecting overlapping sounds from the same class with auxiliary duplicating permutation invariant training},
  author={Shimada, K. and others},
  booktitle={IEEE International Conference on Acoustics, Speech and Signal processing (ICASSP)},
  pages={316--320},
  year={2022},
  organization={IEEE}
}

@inproceedings{kowalk2023geometry,
  title={Geometry-aware DOA estimation using a deep neural network with mixed-data input features},
  author={Kowalk, U. and Doclo, S. and Bitzer, J.},
  booktitle={IEEE International Conference on Acoustics, Speech and Signal Processing (ICASSP)},
  pages={1--5},
  year={2023},
  organization={IEEE}
}

@article{baek2025dnn,
  title={Dnn-based geometry-invariant doa estimation with microphone positional encoding and complexity gradual training},
  author={Baek, M-S. and Chang, J-H. and Cohen, I.},
  journal={IEEE Transactions on Audio, Speech and Language Processing},
  year={2025},
  publisher={IEEE}
}

@article{schmidt1986multiple,
  title={Multiple emitter location and signal parameter estimation},
  author={Schmidt, R.},
  journal={IEEE Transactions on Antennas and Propagation},
  volume={34},
  number={3},
  pages={276--280},
  year={1986},
  publisher={Ieee}
}

@book{dibiase2000high,
  title={A High-Accuracy, Low-Latency Technique for Talker Localization in 
Reverberant Environments Using Microphone Arrays },
  author={DiBiase, J. H.},
  year={2000},
  publisher={Brown University}
}

@article{grumiaux2022survey,
  title={A survey of sound source localization with deep learning methods},
  author={Grumiaux, P.-A. and others},
  journal={The Journal of the Acoustical Society of America},
  volume={152},
  number={1},
  pages={107--151},
  year={2022},
  publisher={AIP Publishing}
}

@article{grinstein2023neural,
  title={The neural-SRP method for universal robust multi-source tracking},
  author={Grinstein, E. and others},
  journal={IEEE Open Journal of Signal Processing},
  volume={5},
  pages={19--28},
  year={2023},
  publisher={IEEE}
}

@inproceedings{schwartz2023array,
  title={Array configuration mismatch in deep DOA estimation: Towards robust training},
  author={Schwartz, A. and others},
  booktitle={IEEE Workshop on Applications of Signal Processing to Audio and Acoustics (WASPAA)},
  pages={1--5},
  year={2023},
  organization={IEEE}
}

@article{dementyev2026phasecoder,
  title={PhaseCoder: Microphone Geometry-Agnostic Spatial Audio Understanding for Multimodal LLMs},
  author={Dementyev, A. and others},
  journal={ArXiv preprint arXiv:2601.21124},
  year={2026}
}

@article{mccormack2023spatial,
  title={Spatial reconstruction-based rendering of microphone array room impulse responses},
  author={McCormack, L. and Meyer-Kahlen, N. and Politis, A.},
  journal={Journal of the Audio Engineering Society},
  volume={71},
  number={5},
  pages={267--280},
  year={2023},
  publisher={Audio Engineering Society}
}

@inproceedings{tachibana2021towards,
  title={Towards listening to 10 people simultaneously: An efficient permutation invariant training of audio source separation using sinkhorn’s algorithm},
  author={Tachibana, H.},
  booktitle={IEEE International Conference on Acoustics, Speech and Signal Processing (ICASSP)},
  pages={491--495},
  year={2021},
  organization={IEEE}
}

@inproceedings{Wichern2019WHAM,
    title={WHAM!: Extending Speech Separation to Noisy Environments},
    author={Wichern, G. and others},
    booktitle = {Proc. Interspeech},
    year      = {2019},
    month     = sep
}

@inproceedings{panayotov2015librispeech,
  title={Librispeech: an asr corpus based on public domain audio books},
  author={Panayotov, V. and others},
  booktitle={IEEE International Conference on Acoustics, Speech and Signal Processing (ICASSP)},
  pages={5206--5210},
  year={2015},
  organization={IEEE}
}

@inproceedings{scheibler2018pyroomacoustics,
  title={Pyroomacoustics: A python package for audio room simulation and array processing algorithms},
  author={Scheibler, R. and Bezzam, E. and Dokmani{\'c}, I.},
  booktitle={IEEE International Conference on Acoustics, Speech and Signal Processing (ICASSP)},
  pages={351--355},
  year={2018},
  organization={IEEE}
}

@article{betcke2021bempp,
  title={Bempp-cl: A fast Python based just-in-time compiling boundary element library},
  author={Betcke, T. and Scroggs, M.},
  journal={Journal of Open Source Software},
  volume={6},
  number={59},
  pages={2879--2879},
  year={2021},
  publisher={The Open Journal}
}

@article{politis2020overview,
  title={Overview and evaluation of sound event localization and detection in DCASE 2019},
  author={Politis, A. and others},
  journal={IEEE/ACM Transactions on Audio, Speech, and Language Processing},
  volume={29},
  pages={684--698},
  year={2020},
  publisher={IEEE}
}

@inproceedings{heikkinen2026beyond,
  title     = {Beyond Omnidirectional: Neural Ambisonics Encoding for Arbitrary Microphone Directivity Patterns Using Cross-Attention},
  author    = {Heikkinen, M. and others},
  booktitle = {Proceedings of the IEEE International Conference on Acoustics, Speech and Signal Processing (ICASSP)},
  year      = {2026},
}

@article{liu2018intriguing,
  title={An intriguing failing of convolutional neural networks and the coordconv solution},
  author={Liu, R. and others},
  journal={Advances in Neural Information Processing Systems},
  volume={31},
  year={2018}
}

@article{vaswani2017attention,
  title={Attention is all you need},
  author={Vaswani, A. and others},
  journal={Advances in Neural Information Processing Systems},
  volume={30},
  year={2017}
}

\end{document}